\documentclass [12pt,a4paper,draft]{article}

\usepackage{times}

\DeclareFontFamily{OT1}{times}{}
\DeclareFontShape {OT1}{times}{m }{n }{ <-> ptmr }{}
\DeclareFontShape {OT1}{times}{bx}{n }{ <-> ptmb }{}
\DeclareFontShape {OT1}{times}{m }{it}{ <-> ptmri}{}
\DeclareFontShape {OT1}{times}{bx}{it}{ <-> ptmbi}{}
\usepackage{amsmath}
\usepackage{amsfonts}
\usepackage{amssymb}
\usepackage{latexsym}
\usepackage{graphics}
\begin{document}

\title{NON-LINEAR  FIELD  THEORY  FOR
LEPTON  AND QUARK MASSES}

\author{André Gsponer and Jean-Pierre Hurni\\
\emph{Independent Scientific Research Institute}\\
\emph{P.O. Box 30, 1211 Geneva--12, Switzerland}\\ 
e-mail: isri@vtx.ch}

\date{Received November 23, 1995}

\maketitle

\begin{abstract}

	Barut's formula for the mass of leptons is successfully 
extended to quarks. A very simple non-linear scalar field model 
explains both the $N^4$ power law dependence of the mass, and the 
existence of a cut-off which limits the number of leptons to three 
and the number of quarks to five, suggesting that the mass of the 
sixth quark is of different origin.

\end{abstract}	

            \begin{center}

      ~

\vspace{2 cm}

Published in ~ {\bf HADRONIC ~ JOURNAL ~ 19} (1996) 367-373

\vspace{2 cm}

{\bf Copyright \copyright 1996 by Hadronic Press, Inc., Palm Harbor, FL 34682, USA}

    \end{center}

\newpage

The foundation of the very successful Standard Model of elementary particles is the concept of partons, i.e. leptons and quarks which interact by means of local gauge fields. The basic characteristics of these partons, such as their mass and charge, are not fully explained by the Model.  Moreover, there is still no explanation of why the top quark weighs much more than its siblings and why a meaningful pattern for the masses of the partons defies understanding.

In this paper we look at the problem of lepton and quark masses from a phenomenological point of view. We therefore take the measured lepton masses, as well as the quark masses (although they are only indirectly measured), as input data and look at regularities. We then give an interpretation of the resulting sequence in terms of non-linear wave mechanics.

Our starting point is the observation that the lepton mass formula discovered by Barut [1] in 1979 can easily be extended to quarks. Assuming that a quantized self-energy of magnitude  $\frac{3}{2} \alpha^{-1} M_e c^2 N^4$, where $\alpha =  \frac{e^2}{\hbar c}$ and  $N = 0, 1, 2, ...$,  is a new quantum number, be added to the rest-mass of a lepton to get the next heavy lepton in the chain $ e, \mu, t, ...$, Barut got the following expression:
$$
  M(N)= M_e (1 + \frac{3}{2} \alpha^{-1} \sum_{n=0}^{n=N} n^4 ) \eqno(1)
$$
The agreement with the data of this  rather simple formula is  surprisingly good, the discrepancy being of the order $10^{-4}$ for $\mu$  and $10^{-3}$ for $\tau$, respectively. In order to get the masses of the quarks, we simply take as the mass of the lightest quark $M_u  = M_e/7.25$ . Leaving the justification of this factor for later, we see in Figure~1 and Table~1 that the agreement between our theoretical quark masses and the "observed" masses [2,3] is quite good.

In his paper [1], Barut suggested that magnetic self-interaction between the charge and the anomalous magnetic moment of a lepton could explain the $N^4$ power law dependence. In fact, as we found out, many different type of mechanisms lead to a quantum number dependence or to a scaling with the fourth power of some variable [4,5]. In particular,  this will  be  the  case  in non-linear  field  theories  with  a $F^4$ term in the Lagrangian [6]. In this letter, we present an example of such a theory.

We postulate that the only stable partons are the electron and the $u$-quark. Let us suppose that all "excited" states of partons can be described by a relativistic bag-like model. The bag is taken as the   "proper sphere" bounded by a surface of constant retarded distance  $s :=  ct - \vec{x}\cdot\vec{v}/c  =  s_0$.  The "mass increment" associated with the $n^{th}$ excitation level is then postulated to be obtained by integrating the energy density of a scalar field over the volume of the proper sphere. Hence the name of the model: the scalar "barybag".

We assume a non-linear Lagrangian density with a $F^4$ term:
$$
	\frac{dL}{dV}=   \frac{1}{2} ~\partial_\nu F ~\partial^\nu F  -  \frac{1}{2}  ~\mu^2 F^2  - \frac{1}{4 g^2} ~F^4  ~~ .		\eqno(2)
$$
As in the traditional bag-model of hadrons [8],  variation of the Lagrangian gives the field equations, and two boundary conditions are needed: The field has to vanish on the boundary, and its gradient has to be orthogonal to the normal to the boundary. The second condition is trivially satisfied if $F = F(\omega s)$ where $\omega$ is a parameter. In order to put our equations in non-dimensional form, we write $F  =  \sqrt{2} \omega g k f(\omega s)$ where $k$ is a pure number.  The field equation is then
$$
 \frac{d^2}{ds^2} f  + \frac{\mu^2}{\omega^2} f + 2 k^2 f^3  =  0  ~~ .	\eqno(3)
$$
The solutions, finite on the real axis, are the Jacobi elliptic functions  $cn$ and $sd$. Their modulus is   $k^2 =  \frac{1}{2} (1 - \frac{\mu^2}{\omega^2})$ and the energy density is found to be independent of $s$ :
$$
    4 \pi \frac{dE}{dV}  =   g^2 \omega^4 k^2 ( 1 - k^2 )   ~~ . \eqno(4)
$$
We now use the boundary condition. The function $cn$ and $sd$ have zeros at arguments of value $(2n+1){\pmb K}$ or $2n{\pmb K}$, where ${\pmb K}(k)$ is the real quarter-period. Hence, the condition that $F$ vanishes on the boundary translates into the quantization condition   $s_0 \omega(n) = n {\pmb K}(k)$.

As the energy density is constant within the barybag, the calculation of the mass increment is trivial. We get
$$
 \Delta M(n) c^2 = \frac{{\pmb K}^4}{3} n^4 \frac{g^2}{s_0}k^2 (1 - k^2) ~~ . \eqno(5)
$$
In the limit  $\mu^4 \ll \omega^4$, which implies  $k^2 = \frac{1}{2}$, we have  ${\pmb K}^4(\surd\frac{1}{2}) \approx 11.81$ . Comparing with Barut's formula (1), or the lepton masses, and postulating that $g$ equals the electric charge quantum $e$, we get $s_0 \approx \frac{2}{3} \alpha r_e$ where  $r_e = \frac{e^2}{M_e c^2}$ is the classical electron radius. Consequently, the barybag radius is equal to the classical electromagnetic radius of the muon, a small fraction of the electromagnetic radius of any hadron.

Our assumption that $g = e$ is consistent with the idea that the elementary particle's basic properties might be explained by a theory in which $\alpha$ is somehow the fundamental interaction constant [9].  Indeed, many correlations involving $\alpha$ have been found by various authors, as much for the particle's 
masses [10] as for their lifetimes [11].

Postulating that $s_0$ is a fundamental length and that $g = e$ for both leptons and quarks, the only adjustable parameter left is $k$.

The values of $k$ which give the lepton and quark mass spectras can be explained in terms of a theory [12] in which the single-periodic "de Broglie waves", that quantum mechanics associates with particles, are generalized to double-periodic "Petiau waves" [13].  Instead of being linear combinations of sin  and  cos  functions, these waves are superpositions of elliptic functions $sn$, $cn$,  etc.  A very appealing feature of Petiau waves is that their dependence on the modulus interpolates between pure de Broglie waves (for $k=0$) and pure solitonic waves (for $k=1$): A beautiful realization of the wave/particle duality of quantum mechanics.

The introduction of Petiau waves implies a non-linear generalization of quantum theory [12]. As shown by Petiau [14], in term of the first integrals, the Hamiltonian of a free particle has the form
$$
		H  =   C_0 \omega^4 k^2 ( 1 - k^2 )   \eqno(6)
$$
where $C_0$ is a constant. This expression has the same form as (4), confirming that the barybag can be interpreted as the envelop of a superposition of Petiau waves confined to a limited space-time region.

There are two non-trivial exceptional cases for elliptic functions: the harmonic case, $k =  \sin(\frac{\pi}{4})$, and the equianharmonic case, $k = \sin(\frac{\pi}{12})$.  It is very plausible to associate the former with leptons, and the latter with quarks. Indeed, in either case, the corresponding elliptic functions exhibit several unique symmetry and scaling properties, which come from the fact that in the complex plane their poles form a modular aggregate with $\frac{\pi}{2}$ or $\frac{\pi}{3}$ symmetries. Using these two special values of $k$ in (5), we obtain a mass ratio of about 7.2448 for the lepton and quark mass sequences.

An important aspect of non-linear quantum theory is that the superposition principle does not apply. There is thus no interference between solutions $F(n,k,s)$ with different values of the quantum number $n$.  This means that the total mass is obtained by simply adding the mass increments given  by (5). Looking at Table~1 and Figure~1, the main discrepancy between our mass formula and the data is the non-existence of partons with sequential masses larger than approximately  $\alpha^{-2}M_e =  9.6$  GeV/c$^2$. This fact suggests the existence of a cut-off.

A simple explanation for the cut-off is provided by Heisenberg's uncertainty principle. In ordinary quantum theory, when a particle is restricted to a region of radius $s_0$, its minimum energy, derived from its momentum uncertainty, is roughly given by
$$
    E_{s_0} \approx \frac{1}{2} \frac{\hbar c}{s_0} = \frac{3}{4} \alpha^{-2} M_e c^2 \approx 7.2 \; {\text GeV} \eqno(7)
$$
Hence, when the energy of the barybag becomes of the order of  $E_{s_0}$, a plausible explanation of the ending of the mass spectrum is that the nonspreading barybag "decays" into a de Broglie wave packet which spreads because it is a superposition of waves of different energies. This interpretation is consistent with Petiau's idea that while "progressive" waves of type $\exp(i\omega s)$ are fundamental in linear quantum theory, "standing" waves of type $cn(\omega s)$  are the fundamental waves in non-linear quantum theory [13].

In the currently Standard Model of elementary particles, a sixth quark is needed for theoretical reasons. The experimental observation of this quark [7] could therefore be interpreted as an argument against the present explanation of the Barut formula. The Standard Model, however, is basically a perturbative theory of the interactions of quarks and leptons, a theory that is logically independent of our theory of mass quantization.

The conjunction of our theory of mass with one of interactions could possibly explain the existence of three very light neutrinos, which have also to be predicted by a complete model,  as well as a very massive sixth quark. A more complete model is also needed to provide a clear-cut distinction between our generalized  $N^4$-Barut formula and alternative mass spectrum formulae such as the one recently proposed by Rosen [5].

A first step in this direction is the observation that the modulus $k^2 = \frac{1}{2}$  implies  $\mu(k) = 0$. Ignoring the non-linear term, the field equation of a massive lepton corresponds therefore to a massless particle: the neutrino. In the quark case however,  $\mu(k) \neq 0$. For  $n = 5$,  $m$  is about 115 GeV/c$^2$, i.e., on the order of the sixth quark mass.

\section{ Table 1}

\begin{tabular}{|c|c|c|c|c|c|c|c|}
\hline
 N & \multicolumn{3} {c|} {electron masses}   & \multicolumn{3} {c|} {quark masses} & \multicolumn{1} {c|} {~} \\
\hline
   &      & Barut's  &  Ref.    &         &  Barut's  &     Ref.     &       Ref.      \\
	
   &      & formula  &  [2]     &         &  formula  &     [2]      &       [3]       \\
\hline
 0 & e    &    0.511 &    0.511 & u       &     0.068 &    0 -- 8    &     8 $\pm$ 2   \\
	
 1 &$\mu$ &  105.55  &  105.66  & d       &    14.1   &    5 -- 15   &    13 $\pm$ 4   \\
	
 2 &$\tau$&  1786.1  & 1784.1   & s       &   239     &  100 -- 300  &   260 $\pm$ 80  \\
	
 3 &      &  10294.  &    ?     & c       &  1378     & 1300 -- 1500 &  1350 $\pm$ 50  \\
	
 4 &      &  37184.  &    ?     & b       &  4978     & 4700 -- 5300 &  5300 $\pm$ 100 \\
	
 5 &      &          &          & t       & 13766     &      ?       &         ?       \\
	
 6 &      &          &          &         & 31989     &      ?       &         ?       \\
\hline
\end{tabular}

Comparison of lepton and quark masses in MeV/c$^2$ calculated with Barut's formula (1) to measured lepton masses from Ref.2 and to quark masses given in Ref.2 and 3. The observation of a sixth quark of mass in the range of 160'000 to 190'000 MeV/c$^2$ has been reported at the beginning of 1995 [7].

\section{ Figure 1}

Theoretical mass of leptons and quarks as a function of sequential quantum number $N$.  Open circles are hypothetical partons with masses larger than 9.6 GeV/c$^2$. The lines through the points are guides for the eyes.  $M_e  = 511$ keV/c$^2$ and $M_u = M_e / 7.25$ are input data. The uncertainty on quark masses are from Ref.2.

\newpage

\setlength{\unitlength}{1mm}
\begin{picture}(120,200)(0,0)
\linethickness{  0.00mm}
\put(0,0){\framebox(120,200){}}
\LARGE
\linethickness{  0.25mm}
\linethickness{  0.50mm}
\put(   20.00,   60.00){\line(  1,  0){  100.00}}
\put(  120.00,   60.00){\line(  0,  1){  100.00}}
\put(  120.00,  160.00){\line( -1,  0){  100.00}}
\put(   20.00,  160.00){\line(  0, -1){  100.00}}
\linethickness{  0.50mm}
\put(   20.00,   60.00){\line(  0,  1){    3.00}}
\put(   20.00,   54.00){\makebox(0,0){   0
}}
\put(   34.29,   60.00){\line(  0,  1){    3.00}}
\put(   34.29,   54.00){\makebox(0,0){   1
}}
\put(   48.57,   60.00){\line(  0,  1){    3.00}}
\put(   48.57,   54.00){\makebox(0,0){   2
}}
\put(   62.86,   60.00){\line(  0,  1){    3.00}}
\put(   62.86,   54.00){\makebox(0,0){   3
}}
\put(   77.14,   60.00){\line(  0,  1){    3.00}}
\put(   77.14,   54.00){\makebox(0,0){   4
}}
\put(   91.43,   60.00){\line(  0,  1){    3.00}}
\put(   91.43,   54.00){\makebox(0,0){   5
}}
\put(  105.71,   60.00){\line(  0,  1){    3.00}}
\put(  105.71,   54.00){\makebox(0,0){   6
}}
\put(  120.00,   60.00){\line(  0,  1){    3.00}}
\put(  120.00,   54.00){\makebox(0,0){   7
}}
\put(  120.00,   60.00){\line( -1,  0){    3.00}}
\put(  120.00,   74.29){\line( -1,  0){    3.00}}
\put(  120.00,   88.57){\line( -1,  0){    3.00}}
\put(  120.00,  102.86){\line( -1,  0){    3.00}}
\put(  120.00,  117.14){\line( -1,  0){    3.00}}
\put(  120.00,  131.43){\line( -1,  0){    3.00}}
\put(  120.00,  145.71){\line( -1,  0){    3.00}}
\put(  120.00,  160.00){\line( -1,  0){    3.00}}
\put(  120.00,  160.00){\line(  0, -1){    3.00}}
\put(  105.71,  160.00){\line(  0, -1){    3.00}}
\put(   91.43,  160.00){\line(  0, -1){    3.00}}
\put(   77.14,  160.00){\line(  0, -1){    3.00}}
\put(   62.86,  160.00){\line(  0, -1){    3.00}}
\put(   48.57,  160.00){\line(  0, -1){    3.00}}
\put(   34.29,  160.00){\line(  0, -1){    3.00}}
\put(   20.00,  160.00){\line(  0, -1){    3.00}}
\put(   20.00,  160.00){\line(  1,  0){    3.00}}
\put(    8.75,  160.00){\makebox(0,0){    $10^5$ }}
\put(   20.00,  145.71){\line(  1,  0){    3.00}}
\put(    8.75,  145.71){\makebox(0,0){    $10^4$ }}
\put(   20.00,  131.43){\line(  1,  0){    3.00}}
\put(    8.75,  131.43){\makebox(0,0){    $10^3$ }}
\put(   20.00,  117.14){\line(  1,  0){    3.00}}
\put(    8.75,  117.14){\makebox(0,0){    $10^2$ }}
\put(   20.00,  102.86){\line(  1,  0){    3.00}}
\put(    8.75,  102.86){\makebox(0,0){    $10^1$ }}
\put(   20.00,   88.57){\line(  1,  0){    3.00}}
\put(    8.75,   88.57){\makebox(0,0){    $10^0$ }}
\put(   20.00,   74.29){\line(  1,  0){    3.00}}
\put(    8.75,   74.29){\makebox(0,0){ $10^{-1}$ }}
\put(   20.00,   60.00){\line(  1,  0){    3.00}}
\put(    8.75,   60.00){\makebox(0,0){ $10^{-2}$ }}
\linethickness{  0.75mm}
\qbezier[200](   20.00,   84.41)(   27.14,  100.94)(   34.29,  117.48)
\qbezier[200](   34.29,  117.48)(   41.43,  126.25)(   48.57,  135.03)
\qbezier[200](   48.57,  135.03)(   55.71,  140.46)(   62.86,  145.89)
\qbezier[200](   62.86,  145.89)(   70.00,  149.88)(   77.14,  153.86)
\qbezier[200](   77.14,  153.86)(   84.29,  157.02)(   91.43,  160.17)
\put(   21.43,   84.41){\makebox(0,0){$\bullet$
}}
\put(   21.43,   84.41){\makebox(0,0){$~~~ e$
}}
\put(   35.71,  117.48){\makebox(0,0){$\bullet$
}}
\put(   35.71,  117.48){\makebox(0,0){$~~ ~~ \mu$
}}
\put(   50.00,  135.03){\makebox(0,0){$\bullet$
}}
\put(   50.00,  135.03){\makebox(0,0){$~~~ ~~~ \tau$
}}
\put(   64.29,  145.89){\makebox(0,0){$\circ$
}}
\put(   78.57,  153.86){\makebox(0,0){$\circ$
}}
\put(   92.86,  160.17){\makebox(0,0){$\circ$
}}
\qbezier[200](   20.00,   71.89)(   27.14,   88.45)(   34.29,  105.00)
\qbezier[200](   34.29,  105.00)(   41.43,  113.78)(   48.57,  122.55)
\qbezier[200](   48.57,  122.55)(   55.71,  127.98)(   62.86,  133.42)
\qbezier[200](   62.86,  133.42)(   70.00,  137.40)(   77.14,  141.39)
\qbezier[200](   77.14,  141.39)(   84.29,  144.54)(   91.43,  147.70)
\qbezier[200](   91.43,  147.70)(   98.57,  150.31)(  105.71,  152.93)
\qbezier[200](  105.71,  152.93)(  112.86,  155.16)(  120.00,  157.40)
\put(   21.43,   71.89){\makebox(0,0){$\bullet$
}}
\put(   21.43,   71.89){\makebox(0,0){$~~ ~~ u$
}}
\put(   35.71,  105.00){\makebox(0,0){$\bullet$
}}
\put(   35.71,  105.00){\makebox(0,0){$~~ ~~ d$
}}
\put(   50.00,  122.55){\makebox(0,0){$\bullet$
}}
\put(   50.00,  122.55){\makebox(0,0){$~~ ~~ s$
}}
\put(   64.29,  133.42){\makebox(0,0){$\bullet$
}}
\put(   64.29,  133.42){\makebox(0,0){$~~ ~~ ~~ c$
}}
\put(   78.57,  141.39){\makebox(0,0){$\bullet$
}}
\put(   78.57,  141.39){\makebox(0,0){$~~ ~~ ~~ b$
}}
\put(   92.86,  147.70){\makebox(0,0){$\circ$
}}
\put(  107.14,  152.93){\makebox(0,0){$\circ$
}}
\put(  121.43,  157.40){\makebox(0,0){$\circ$
}}
\qbezier[200](   48.57,  144.29)(   50.00,  144.29)(   51.43,  144.29)
\qbezier[200](   54.29,  144.29)(   55.71,  144.29)(   57.14,  144.29)
\qbezier[200](   60.00,  144.29)(   61.43,  144.29)(   62.86,  144.29)
\qbezier[200](   65.71,  144.29)(   67.14,  144.29)(   68.57,  144.29)
\qbezier[200](   71.43,  144.29)(   72.86,  144.29)(   74.29,  144.29)
\qbezier[200](   77.14,  144.29)(   78.57,  144.29)(   80.00,  144.29)
\qbezier[200](   82.86,  144.29)(   84.29,  144.29)(   85.71,  144.29)
\qbezier[200](   88.57,  144.29)(   90.00,  144.29)(   91.43,  144.29)
\qbezier[200](   94.29,  144.29)(   95.71,  144.29)(   97.14,  144.29)
\qbezier[200](  100.00,  144.29)(  101.43,  144.29)(  102.86,  144.29)
\put(   37.14,  144.29){\makebox(0,0){$\frac{M_e}{\alpha^2}$
}}
\put(   77.14,  110.00){\makebox(0,0){mass
}}
\put(   77.14,   95.71){\makebox(0,0){of ~ leptons
}}
\put(   77.14,   81.43){\makebox(0,0){and ~ quarks
}}
\put(   70.00,   38.57){\makebox(0,0){$N$
}}
\put(   70.00,    2.86){\makebox(0,0){Fig.1
}}
\put(   -8.57,  110.00){\makebox(0,0){\rotatebox{ 90}{$M$ [MeV/c$^2$]}
}}
\end{picture}

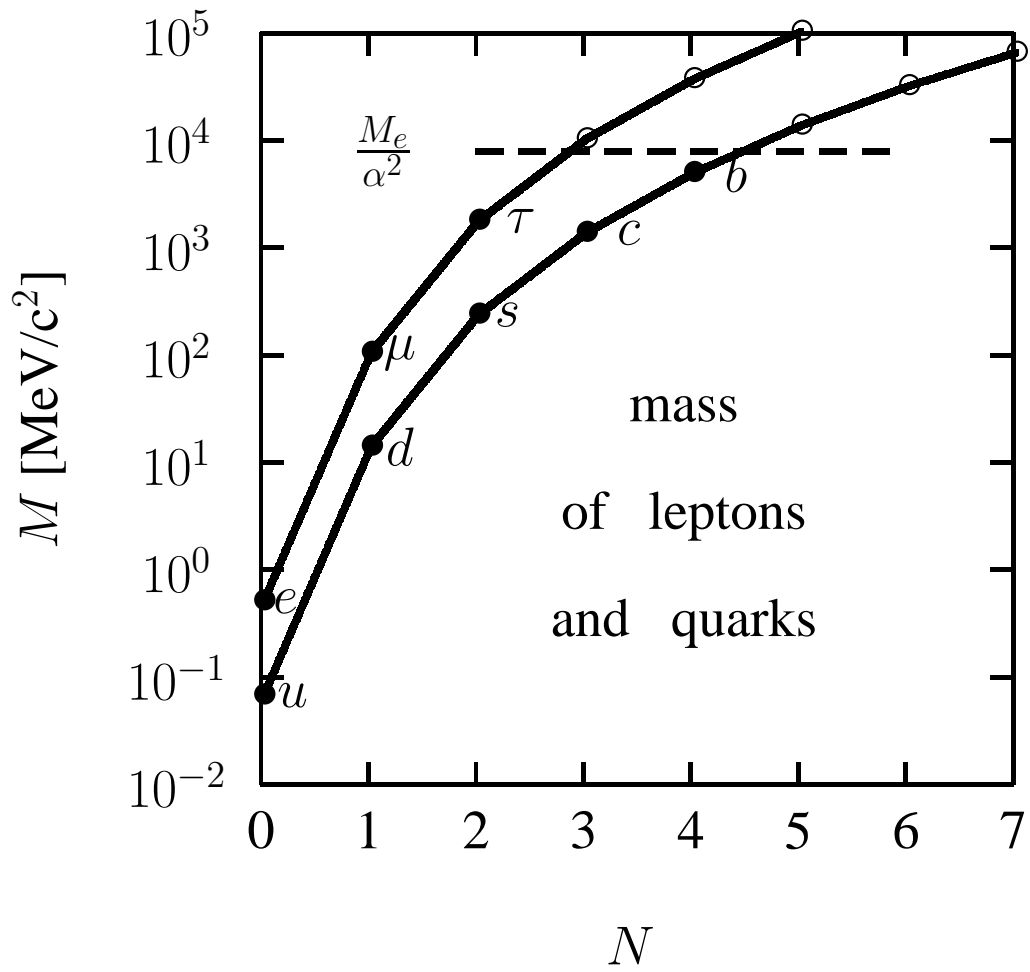
\end{document}